\documentclass[superscriptaddress,twocolumn,showpacs,preprintnumbers,amsmath,amssymb,prb]{revtex4}

\usepackage{graphicx}
\usepackage{dcolumn}
\usepackage{bm}

\usepackage{hyperref} 
\usepackage{hypernat}

\usepackage{color} 
\definecolor{rltred}{rgb}{0.75,0,0}
\definecolor{rltgreen}{rgb}{0,0.6,0}
\definecolor{rltblue}{rgb}{0.3,0.3,1}

\hypersetup{colorlinks,%
    hypertexnames=true,%
    linkcolor=rltgreen,%
    citecolor=rltblue,%
    linktocpage=true}
    
\begin{document}
\title{Wave chaos in the non-equilibrium dynamics of the Gross-Pitaevskii equation}

\author{Iva B\v rezinov\'a}
\email{iva.brezinova@tuwien.ac.at}
\affiliation{Institute for Theoretical Physics, Vienna University of Technology,
Wiedner Hauptstra\ss e 8-10/136, 1040 Vienna, Austria, EU}  

\author{Lee Collins}
\affiliation{Theoretical Division, Los Alamos National Laboratory, Los Alamos, New Mexico 87545, USA}

\author{Katharina Ludwig}
\affiliation{Institute for Theoretical Physics, Vienna University of Technology,
Wiedner Hauptstra\ss e 8-10/136, 1040 Vienna, Austria, EU}

\author{Barry Schneider}
\affiliation{Physics Division, National Science Foundation, Arlington, Virginia 22230, USA}
\affiliation{Electron and Atomic Physics Division, National Institute of Standards and Technology, Gaithersburg, Maryland 20899, USA}

\author{Joachim Burgd\"orfer}
\affiliation{Institute for Theoretical Physics, Vienna University of Technology,
Wiedner Hauptstra\ss e 8-10/136, 1040 Vienna, Austria, EU}  

\date{\today}

\begin{abstract}
The Gross-Pitaevskii equation (GPE) plays an important role in the description of Bose-Einstein condensates (BECs) at the mean-field level. The GPE belongs to the class of non-linear Schr\"odinger equations which are known to feature dynamical instability and collapse for attractive non-linear interactions. We show that the GPE with repulsive non-linear interactions typical for BECs features chaotic wave dynamics. We find positive Lyapunov exponents for BECs expanding in periodic and aperiodic smooth external potentials as well as disorder potentials. Our analysis demonstrates that wave chaos characterized by the exponential divergence of nearby initial wavefunctions is to be distinguished from the notion of non-integrability of non-linear wave equations. We discuss the implications of these observations for the limits of applicability of the GPE, the problem of Anderson localization, and the properties of the underlying many-body dynamics.
\end{abstract}
\maketitle
\section{Introduction}\label{sec:intro}
Following the experimental realization of Bose-Einstein condensates (BECs) in dilute ultracold gases, the Gross-Pitaevskii equation (GPE), has taken center stage to describe the equilibrium as well as non-equilibrium dynamics of the condensate at the mean-field level\cite{proceed98}. The replacement of the many-body wavefunction by the effective single-particle condensate wavefunction has proven to be a remarkably successful approximation for predicting a large variety of physical observables.\cite{DalGioPitStr99} Among the observables are both ground-state properties and elementary excitations in inhomogeneous background potentials.\cite{DalGioPitStr99,LyeFalModWieForIng05,ForFalGuaLyeModWieIng05,SchDreKruErtArlSacZakLew05,CleVarHugAsp05,BilJosZuoCleSanBouyAsp08,Inguscio08,DriPolHitHul10} The GPE belongs to the class of non-linear Schr\"odinger equations (NLSEs) which have a broad range of applications ranging from nonlinear optics to plasma physics and Bose-Einstein condensation.\cite{DauPey06} Effects beyond the GPE have been observed in BECs most notably in optical lattices with deep wells and small occupation numbers per site. In this regime, explicit many-body descriptions such as the Bose-Hubbard model are more suitable.\cite{JakZol04,BloDalZwe08}\\
The non-equilibrium dynamics of BECs, specifically their expansion in disordered potentials has recently received a lot of attention (see e.~g.~Ref.~\onlinecite{LyeFalModWieForIng05,ForFalGuaLyeModWieIng05,SchDreKruErtArlSacZakLew05,CleVarHugAsp05,BilJosZuoCleSanBouyAsp08,Inguscio08,CheHitJunWelHul08,DeiZacRoaErrFatModModIng10,SanLew10,DriPolHitHul10,PauLebPavRicSch05,SanCleLugBouShlAsp07,AlbPauPavLeb08,SkiMinTigSha08,FlaKriSko09,ErnPauSch10} and references therein). One focus is on the observation of Anderson localization of a quantum gas. For weak disorder potentials $\langle V \rangle\ll\mu$, where $\langle V \rangle$ is the variance of the potential and $\mu$ is the chemical potential of the BEC, the GPE was assumed to be valid during the non-equilibrium expansion starting from the BEC released from the trap to the dilute localized state for which the linear one-particle Schr\"odinger limit is reached.\cite{SanCleLugBouShlAsp07,SanLew10} The consequences of the presence of the non-linearity for the non-equilibrium dynamics in a disordered potential described by the GPE deserves a careful analysis. The non-linear Schr\"odinger equation with attractive interactions is known to feature dynamical instabilities leading to collapse of the wavepacket.\cite{DalGioPitStr99} Closely related, the GPE with repulsive pair interaction in a strictly periodic potential features near the Brillouin zone boundary a dynamical (modulation) instability since the effective negative-mass dispersion translates into an effective attractive pair interaction (see e.~g.~Ref.~\onlinecite{WuNiu01,WuNiu03,KonSal02,MacPetSmi03,MorObe06} and references therein). Discretized models resembling the Fermi-Pasta-Ulam-Tsingou\footnote{The model is in literature known under the name Fermi-Pasta-Ulam.\cite{FerPasUla55} We follow here the suggestion by T.~Dauxois\cite{Dau08} to recognize the important contribution of M.~Tsingou to this pioneering computational study.} system of non-linearly coupled oscillators have been found to feature stochastic dynamics and relaxation with an increasing entropy.\cite{HerAbl89,VilLew00,CasMasDunOls09}\\
We show in the following that the GPE for realistic parameters for the expansion of BECs in the quasi-one-dimensional (1D) regime displays true wave chaos as measured by a positive Lyapunov exponent in Hilbert space. By careful checks of the accuracy of the propagation including the method of time-reversed propagation, this ``physical'' chaos can be distinguished from the numerical chaos previously observed for the NLSE.\cite{HerAbl89,AblSchHer93} We furthermore show that chaos goes beyond the non-integrability of non-linear wave equations. The physical consequences of deterministic chaos in the GPE for smooth periodic and aperiodic potentials as well as disorder potentials will be discussed. We argue that wave chaos in the GPE is a signature for the breakdown of mean-field theory and delimits the border of its applicability. The latter does not preclude that certain ensemble expectation values of a BEC can be approximately accounted for by a GPE. We conjecture that the chaotic fluctuations are a signature of excitations and depletion of the condensate. Although our physical interpretations focus on BECs, our findings are relevant for other areas of application of the NLSE as well.\cite{LahAviPozSorMorChrSil08}\\
The paper is organized as follows. In Sec.~\ref{sec:gpe} we briefly describe the model for the expansion of a quasi-1D BEC within the GPE. Numerical methods for the propagation of the condensate wavefunctions will be reviewed in Sec.~\ref{sec:num}. Non-integrability and wave chaos will be discussed in Sec.~\ref{sec:integ} followed by numerical results in Sec.~\ref{sec:examp}. Conclusions and conjectures will be given in Sec.~\ref{sec:dis}. 
\section{Gross-Pitaevskii equation for an expanding (quasi)-1D BEC}\label{sec:gpe}
The GPE for the inhomogeneous condensate wavefunction $\psi(\vec{r},t)=\langle \hat{\psi}(\vec{r},t)\rangle$, the non-vanishing expectation value of the field operator $\hat \psi(\vec{r},t)$ that becomes finite at the Bose-Einstein phase transition, is given by 
\begin{equation}\label{eq:gpe3d}
i\hbar\frac{\partial \psi(\vec{r},t)}{\partial t}=\left(-\frac{\hbar^2\nabla^2}{2m}+V(\vec{r})+g_{\rm{3D}}|\psi(\vec{r},t)|^2\right)\psi(\vec{r},t).
\end{equation}
The effective inter-particle non-linear coupling constant in three dimensions,
\begin{equation}
g_{\rm{3D}}=\frac{4\pi\hbar^2 a_{\rm s}}{m},
\end{equation}
is expected to account for the condensate dynamics under the conditions of low excitations, i.e. weak depletion of the condensate, and of weak coupling. Moreover, Eq.~\ref{eq:gpe3d} assumes short-range interactions at low energies such that the particle-particle interaction can be described by a contact interaction whose strength is proportional to the s-wave scattering length $a_{\rm s}$. The mean-field approximation is assumed to be valid in the dilute regime $n^{1/3}a_{\rm s}\ll1$, where $n$ is the particle density. We consider in the following a (quasi)-1D system, for example, a cigar-shaped trap where the radial and longitudinal frequencies are related as $\omega_{\rm r}\gg\omega_{l}$ and we assume the chemical potential $\mu$ to be small compared to the transverse quantization energy $\mu\ll\hbar\omega_r$.
Thus, the BEC is described by a 1D order parameter. In the transverse direction the dynamics is confined to the groundstate. The 1D GPE is then given by 
\begin{eqnarray}
i\hbar\frac{\partial \psi(x,t)}{\partial t}=-\frac{\hbar^2}{2m}\frac{\partial^2}{\partial x^2}\psi(x,t)+V(x)\psi(x,t)\nonumber \\
+2\hbar\omega_r a_{\rm s}N|\psi(x,t)|^2\psi(x,t).
\end{eqnarray}
The number of atoms $N$ enters explicitly because we normalize the order parameter (in the following termed wavefunction) as $\int \rm{d}x |\psi(x,t)|^2=1$. The potential $V(x)$ is an external potential to be described in more detail below. Measuring energies in units of the longitudinal oscillator energy $\hbar \omega_{l}$, length in units of the oscillator length $l_0=(\hbar/m\omega_{l})^{1/2}$ and time in units $t_0=1/\omega_{l}$ the GPE takes in these (oscillator) units the form
\begin{eqnarray}
i \frac{\partial \psi(x,t)}{\partial t}=
-\frac{1}{2}\frac{\partial^2 \psi(x,t)}{\partial x^2}+U(x)\psi(x,t) \nonumber \\ 
+g|\psi(x,t)|^2\psi(x,t),
\label{eq:gpe_scal}
\end{eqnarray}
with $U(x)=V(xl_0)/\hbar\omega_{l}$, $g=2(\omega_{\rm r}/\omega_{l})aN$ and $a=a_{\rm s}/l_0$. In the following we refer to the effective Hamiltonian of the linear Schr\"odinger equation as $H_L=-\frac{1}{2}\frac{\partial^2}{{\partial}x^2}+U(x)$ and to the non-linear part of the Hamiltonian as $H_{\rm NL}=g|\psi|^2$.\\
Most of the numerical simulations presented in the following are performed for ultracold gases of Rb$^{87}$ initially stored in a cigar-shaped trap with the following parameters: $N=1.2\times 10^4$ and $a_{\rm s}=5.82 {\rm nm}$, and following Ref.~\onlinecite{BilJosZuoCleSanBouyAsp08} $\omega_{l}=5.4\times 2\pi\ \mathrm{Hz}$ and $\omega_{\rm r}=70\times 2\pi\ \mathrm{Hz}$. Accordingly, the unit of time is $t_0=29.47 {\rm ms}$ and the unit of length is $l_0=4.64 {\rm \mu m}$. The associated non-linearity is
\begin{equation} 
\label{eq:g0}
g_0=2\frac{\omega_r}{\omega_l}aN\approx390
\end{equation}
for later reference. Note that the rather high numerical value for $g_0$ is due to the explicit inclusion of the number of particles $N$ and does not contradict the assumption of weak coupling. For these parameters $\mu\ll\hbar\omega_r$ is not fulfilled, however, it is assumed that the BEC is in the quasi-1D regime such that a 1D order parameter is sufficient to describe the dynamics (see e.g.~Ref.~\onlinecite{CleVarHugAsp05,SanCleLugBouShlAsp07,BilJosZuoCleSanBouyAsp08,AlbPauPavLeb08} where the dynamics is essentially confined to its groundstate in the transverse direction).\\
The GPE (Eq.~\ref{eq:gpe_scal}) can be derived as the Euler-Lagrange equation from a Lagrangian functional in the fields $\psi(x,t)$ and $\dot{\psi}(x,t)$. The corresponding Hamiltonian functional is given by 
\begin{equation}
H[\psi]=\int\mathrm{d}x \left(\frac{1}{2}|\partial_x\psi|^2+U|\psi|^2+\frac{g}{2}|\psi|^4\right).
\label{eq:efunc}
\end{equation}
The total energy functional $E=H[\psi]$ must be conserved during the expansion of the BEC. This conservation law provides a stringent test for numerical stability of the long-time propagation.
\section{Numerical methods}\label{sec:num}
The propagation of an initial condensate wavefunction $\psi(x,t=0)$ according to the GPE (Eq.~\ref{eq:gpe_scal}) proceeds by discretization of space and time. The space discretization must be performed carefully since simple discretization schemes may unintentionally convert the integrable continuous GPE into a discrete non-linear system that displays chaos. For example, using the simplest finite difference scheme with $\delta x=x_{\rm i+1}-x_{\rm i}$, Eq.~\ref{eq:gpe_scal} with $U=0$, takes the form 
\begin{eqnarray}
i\partial_t\psi_i=-\frac{1}{2\delta x^2}(\psi_{i+1}-2\psi_i+\psi_{i-1})+g|\psi_i|^2\psi_i.
\label{eq:gpe_fd}
\end{eqnarray}
Eq.~\ref{eq:gpe_fd} shows chaos\cite{HerAbl89} and has been used to study stochastic dynamics and thermalization in the mean-field Bose-Hubbard system\cite{CasMasDunOls09} while its continuous limit is known to be integrable.\cite{HerAbl89} Since we want to study the continuous system we have to avoid such discretization artifacts by employing a more elaborate discretization. Our spatial discretization is based on the finite element discrete variable representation (FEDVR) (see Ref.~\onlinecite{ResCur00,McCHorRes01,SchCol05,SchColHu06} and references therein). We split the space into finite elements which are discretized with the help of a discrete variable representation (DVR) basis. The basis consists of Lagrange interpolating polynomials determined via a grid consisting of the zeros of Legendre polynomials and the endpoints of each element. The elements are connected via bridge functions, thus guaranteeing the continuity of the wavefunctions. Integrals are approximated via the Gauss-Lobatto quadrature. In the FEDVR the local operators (e.~g.~the potential operators) are diagonal. The kinetic operator within a single element is a full $N_\mathrm{b}\times N_\mathrm{b}$ matrix when $N_\mathrm{b}$ is the number of basis functions. Within the FEDVR the Hamiltonian is a sparse matrix when the number of elements and basis functions are chosen appropriately.\\
For temporal propagation we tested several different propagators: the second-order difference\cite{askcak78,LefBisCer91} (SOD), Runge-Kutta\cite{EngUhl96}, the real space product split operator\cite{DeR87,LefBisCer91} (equivalent to the symplectic operator\cite{LasRob01} SABA$_1$), Crank-Nicholson\cite{askcak78} and Lanczos propagators.\cite{parlig86,LefBisCer91} The Runge-Kutta and SOD algorithms have proven to be most accurate in terms of energy conservation. In most of our calculations we use the SOD algorithm. It is an explicit conditionally stable integration scheme given by the recursion relation:
\begin{equation}
\psi(x,t+\mathrm{d}t)=\psi(x,t-\mathrm{d}t)-i2\mathrm{d}t\hat{H}\psi(x,t),
\end{equation}
where $\hat{H}$ is given by
\begin{equation}
\hat{H}=-\frac{1}{2}\frac{\partial^2}{\partial x^2}+U(x)+g|\psi(x,t)|^2.
\end{equation}
The numerically more expensive 4$^{\text{th}}$ order Runge-Kutta algorithm is used for cross-checking. Comparison with a 5$^{\text{th}}$ order Runge-Kutta algorithm allows an accuracy estimate in $|\psi|^2$ to be of the order of $10^{-15}$ for each time step.\\
Both SOD and Runge-Kutta are not symplectic. The Hamiltonian structure of the GPE would suggest to use a symplectic integrator. Such integrators preserve, by design, the volume in phase space which is characteristic for the dynamics of Hamiltonian systems. However, it has been demonstrated for the NLSE and the Korteweg-deVries equation and conjectured for infinite dimensional Hamiltonian systems in general that the use of symplectic time integrators is less important than high accuracy of spatial derivatives.\cite{AblHerSch97,KouZhaWeiKonHof99} For the results presented in the following we use the time-increment ${\rm d}t=4\times10^{-6}$ and $5$ basis functions for finite-element widths of typically $\text{d}x=0.08$ in a box of size $x\in[-1000,1000]$. (For the long time propagations presented in Sec.~\ref{subsec:wd} the box size is $x\in[-1600,1600]$.) We use hard wall boundary conditions for our numerical box. The numerical box is chosen sufficiently large such that the spreading wavepacket $\psi(x,t)$ does not effectively reach the walls of the box within the time of propagation.\\
Apart from energy conservation, an additional sensitive test of the numerical stability of the propagation algorithm is the time reversal of propagation. Expressing the propagation of the initial condensate wavefunction $\psi(x,0)$ as 
\begin{equation}
\psi(x,t)=U_t[\psi(x,0)],
\end{equation}
the time reversed propagation 
\begin{equation}
\psi(x,0)=U_{-t}[\psi(x,t)]
\label{eq:bprop}
\end{equation}
should recover the initial state. For the linear Schr\"odinger equation (LSE), reaching the initial state via Eq.~\ref{eq:bprop} is easier than for the NLSE due to the intrinsic stability of the linear dynamical evolution. For the NLSE, Eq.~\ref{eq:bprop} provides a measure for the accumulation of numerical noise, i.~e.~a measure for ``numerical chaos''.\cite{AblSchHer93} In order to distinguish true deterministic (``physical'') chaos from numerical noise, we will probe the accuracy of Eq.~\ref{eq:bprop} by quantifying how close the backward propagated wavefunction will be to the initial state $\psi(x,0)$.\\
Another stringent test results fom the scaling property of the GPE. We subject the space and time coordinates to the scaling ($\gamma >0$)
\begin{eqnarray}
t\rightarrow \bar{t}&=&t/\gamma \nonumber \\
x\rightarrow \bar{x}&=&x/\sqrt{\gamma}.
\end{eqnarray}
The GPE becomes, after multiplication by $\gamma$,
\begin{eqnarray}
i\frac{\partial}{\partial \bar{t}}\psi(\bar{x},\bar{t})&=&-\frac{1}{2}\frac{\partial^2}{\partial \bar{x}^2}\psi(\bar{x},\bar{t})+\bar{U}(\bar{x})\psi(\bar{x},\bar{t})\nonumber \\
&+&\gamma g |\psi(\bar{x},\bar{t})|^2\psi(\bar{x},\bar{t}),
\label{eq:gpe_rescal1}
\end{eqnarray}
with $\bar{U}(\bar x)=\gamma U(\sqrt{\gamma}x)$. The normalization of the condensate wavefunction after rescaling 
\begin{equation}
\bar\psi(\bar x,\bar t)=\gamma^{1/4}\psi(\bar x,\bar t)
\end{equation}
leads to the rescaled GPE
\begin{eqnarray}
i\frac{\partial}{\partial \bar{t}}\bar{\psi}(\bar{x},\bar{t})&=&-\frac{1}{2}\frac{\partial^2}{\partial \bar{x}^2}\bar{\psi}(\bar{x},\bar{t})+\bar{U}(\bar{x})\bar{\psi}(\bar{x},\bar{t})\nonumber \\
&+&\bar g(\gamma) |\bar{\psi}(\bar{x},\bar{t})|^2\bar{\psi}(\bar{x},\bar{t}).
\label{eq:gpe_rescal2}
\end{eqnarray}
with $\bar g(\gamma)=\sqrt{\gamma}g$. Converged numerical solutions that are not subject to numerical noise will satisfy the scaling behavior (Eq.~\ref{eq:gpe_rescal2}) for every positive value of $\gamma$. Scaling is equivalent to a continuous variation of the spatio-temporal grid. One further consequence of Eq.~\ref{eq:gpe_rescal2} is that scaling corresponds to free tuning of the non-linearity in the absence of $\bar U$. Therefore, thresholds for critical strength of non-linearity, if they exist at all, can occur only in the presence of external potentials.
\section{Integrability, non-integrability and wave chaos in the GPE}\label{sec:integ}
\begin{figure}[tb]
	\centering
		\includegraphics[width=8cm]{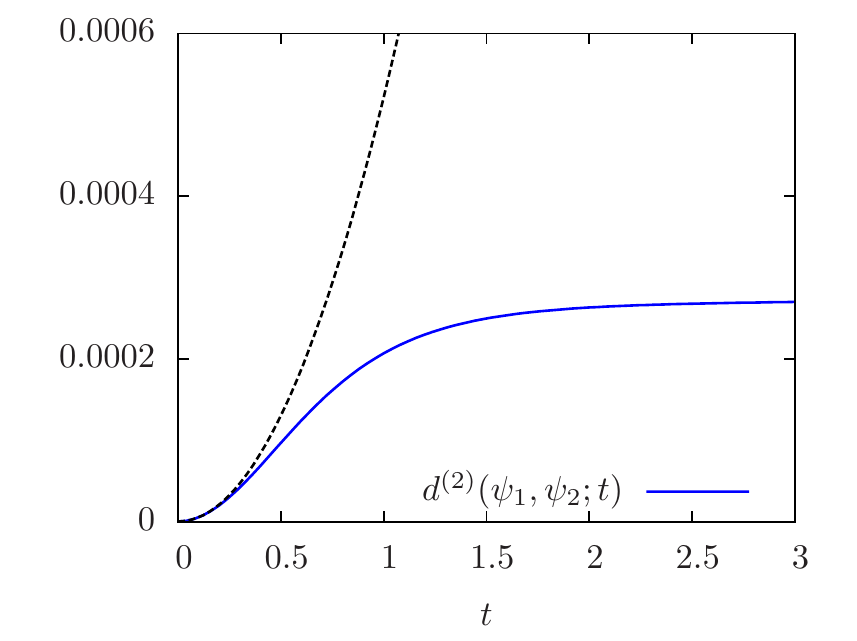}
	\caption{(Color online) Distance function $d^{(2)}(\psi_1,\psi_2;t)$ (Eq.~\ref{eq:d2}) for a freely expanding BEC ($U(x)=0$) with non-linearity $g_0$ released from a harmonic trap (blue solid line). The initial normalized states $\psi_1$, $\psi_2$ are slightly perturbed ground states of the interacting system in the harmonic trap with a small linear perturbation (same states as will be used and explained in more detail in Eq.~\ref{eq:init_wfn} in Sec.~\ref{sec:examp}). The initial distance function is $d^{(2)}(0)\approx2.8\times10^{-7}$. The black dashed line corresponds to $d^{(2)}(\psi_1,\psi_2;t)$ according to Eq.~\ref{eq:d2_2} and Eq.~\ref{eq:re}.}
	\label{fig:integ_d2}
\end{figure}

\begin{figure}[tb]
	\centering
		\includegraphics[width=8.5cm]{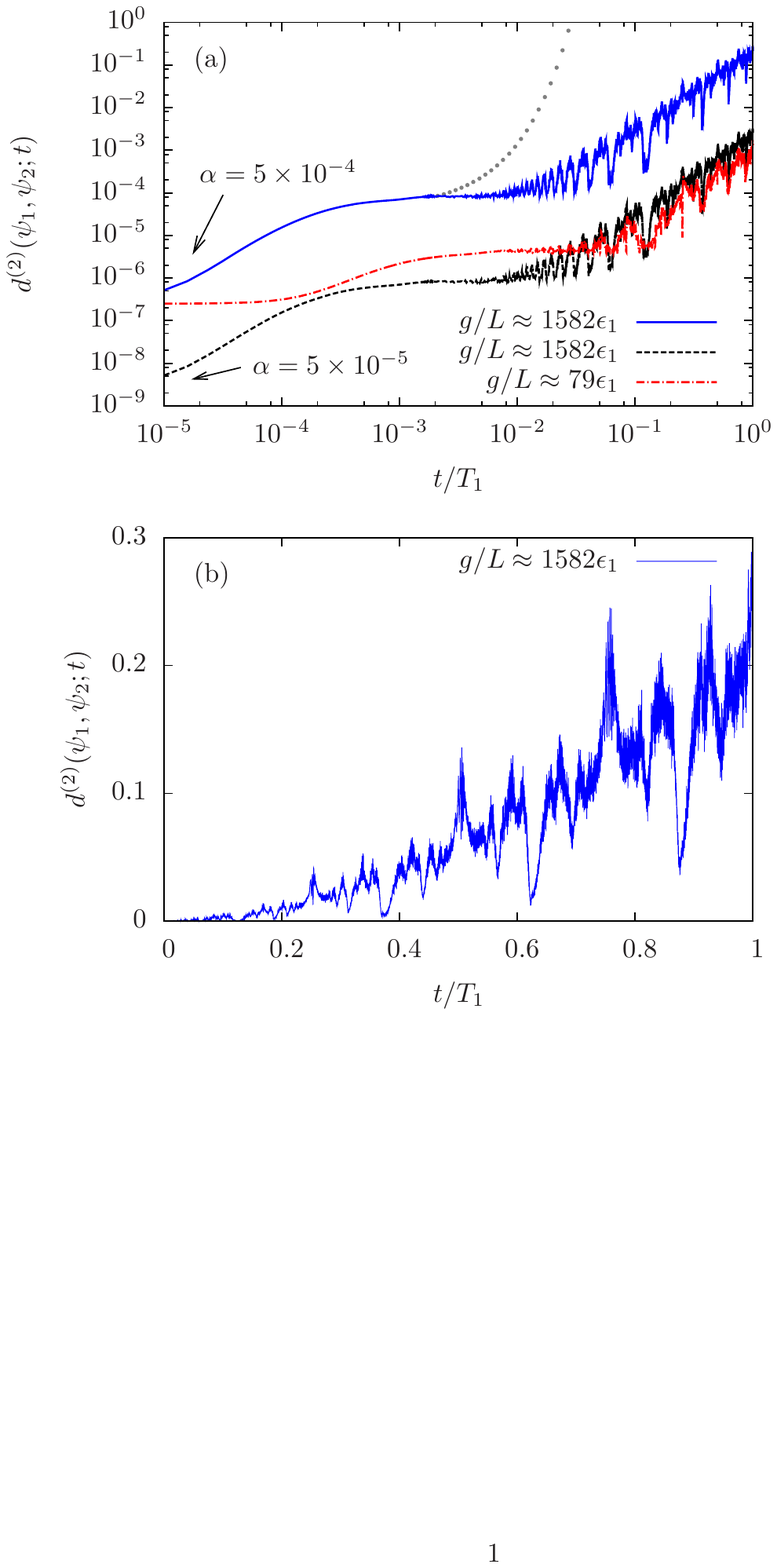}
	\caption{(Color online) (a) Log-log plot of the distance function $d^{(2)}(\psi_1,\psi_2;t)$ (Eq.~\ref{eq:d2}) as a function of $t$ for two initially nearby Gaussian wavepackets in the box of length $L$ obtained from propagation of the GPE with moderate non-linearity ($g/L\approx79\epsilon_1$) and strong non-linearity ($g/L\approx1582\epsilon_1$). For $g/L\approx1582\epsilon_1$ two different distortion parameters $\alpha$ (see Eq.~\ref{eq:init_wfn_gauss}) were used, $\alpha=5\times10^{-4}$ and $\alpha=5\times10^{-5}$ (marked by arrows in the figure). Time is measured in units of $T_1=2\pi/\epsilon_1$. For reference, the exponential growth in case of wave chaos with a Lyapunov exponent as in Fig.~\ref{fig:perlat_d2_fb} is shown as dotted line. (b) The curve with $g/L\approx1582\epsilon_1$ and $\alpha=5\times10^{-4}$ in linear scale.}
	\label{fig:box_d2}
\end{figure}

For non-linear wave equations, the GPE being an example of which, integrability is closely associated with the existence of solutions via the inverse scattering transform (IST). Briefly, the initial value problem (IVP) of the non-linear wave equation possesses an integrable solution if the solution can be determined by IST. In such a case, the non-linear wave equation is associated with a system of auxiliary linear ordinary differential equations (LODEs) for which $\psi(x,t)$ acts as a ``potential'' and whose direct and inverse scattering solution solves the original IVP (see e.~g.~Ref.~\onlinecite{EckHar81,AblCla91,DraJoh96}). The time $t$ acts here as a ``deformation parameter''. While this criterion for integrability is explicit and constructive, it does not, however, provide a priori a sufficient criterion for non-integrability.\\
For the GPE, a few special cases are known: For vanishing external potential, $U(x)=0$, and infinite domain ($-\infty <x<\infty$), the GPE is integrable and the corresponding LODEs are the Zakharov-Shabat (ZS) system.\cite{EckHar81,AblCla91,DraJoh96} While for attractive inter-particle interactions ($g<0$), the ZS system features both a continuum and bound states corresponding to soliton solutions of the GPE, for repulsive interactions ($g>0$), the case of interest in the following, the ZS spectrum is purely continuous corresponding to a dispersing (decaying) condensate wavefunction $\psi(x,t)$. For hard-wall Dirichlet boundary conditions, the GPE with $U(x)=0$ was conjectured to be non-integrable.\cite{VilLew00}\\
For non-vanishing external potentials ($U(x)\ne0$), no general results on the existence of integrable solutions of the GPE are known. For a few special cases, integrability has been demonstrated. These cases include the harmonic potential, $U(x)\propto x^2$, with  time-dependent non-linearities,\cite{LiaZhaLiu05,AtrPanAga06,SerHasBel07} and linear potentials, $U(x)\propto x$, with time-independent non-linearity.\cite{CheLiu76,Kha07}\\
One test for integrability is the Weiss-Tabor-Carnevale (WTC) test for non-autonomous NLSEs with time and space dependent dispersion, non-linearity and dissipation.\cite{AblCla91} The WTC test is based on the Painlev\'e conjecture (see e.g.~Ref.~\onlinecite{AblCla91,DraJoh96}) and provides a necessary condition for integrability (see Ref.~\onlinecite{HeZhaLiLuo09} for specific potentials and nonlinearities). Another integrability condition is based on the Lax pair method yielding, in general, different criteria.\cite{Kha10} However, both integrability conditions agree in that for constant dispersion and nonlinearity and vanishing dissipation the potential $U(x)$ may be at most linear in $x$. For disordered potentials in the NLSE it was shown that quasi-periodic motion may persist for weak non-linearities that can be considered as a small perturbation.\cite{Kuk93}\\
In the following we consider potentials $U(x)$ of physical interest which do not satisfy the known criteria of integrability. They include disordered, periodic and aperiodic potentials. We will characterize the resulting properties of the GPE by a stronger criterion than non-integrability, i.~e.~the appearance of wave chaos. As a measure for wave chaos we employ the existence of a positive Lyapunov exponent which indicates exponential sensitivity to initial conditions. In analogy to classical (particle) chaos, where the distance of initially close trajectories is followed in phase space, we follow the distance of initially close wavefunctions in Hilbert space. The metric in Hilbert space is the L$^2$norm. Accordingly, we use a distance function
\begin{eqnarray}
\label{eq:d2}
d^{(2)}(\psi_1,\psi_2;t)&=&\frac{1}{2}\langle\psi_1-\psi_2|\psi_1-\psi_2\rangle\\ \nonumber
&=&\frac{1}{2}\int_{-\infty}^{\infty} {\rm d}x\;|\psi_1(x,t)-\psi_2(x,t)|^2.
\end{eqnarray}
The Lyapunov exponent follows as the limit
\begin{equation}
\label{eq:lya}
\lambda=\frac{1}{2}\lim_{t\rightarrow\infty}\lim_{d^{(2)}(\psi_1,\psi_2;0)\rightarrow0}\ln{\frac{d^{(2)}(\psi_1,\psi_2;t)}{d^{(2)}(\psi_1,\psi_2;0)}}.
\end{equation}
In analogy to chaos for point particles, $\lambda$ is only well-defined in terms of a coupled double limit. The distance function $d^{(2)}$ (Eq.~\ref{eq:d2}) for normalized functions and norm-conserving evolution reduces to 
\begin{equation}
\label{eq:d2_2}
d^{(2)}(\psi_1,\psi_2)=1-\operatorname{Re}{\langle\psi_1|\psi_2\rangle}.
\end{equation}
It is bounded by $0\leq d^{(2)}\leq 2$. The upper bound corresponds to $\psi_1=-\psi_2$. The value $1$ is reached for complete orthogonality and thus corresponds to the ``maximal'' separation in Hilbert space. Therefore, we expect that an exponential separation of initially ``nearby'', i.~e.~almost identical wavefunctions will eventually saturate at values $d^{(2)}\approx 1$. Eq.~\ref{eq:lya} can be viewed as a measure for the exponential separation of trajectories in Hilbert space on the hypersphere $S$ of unit radius defined by wavefunctions of unit norm. In practice, $\lambda$ is numerically determined by the slope of that segment of growth on a semi-log plot that is approximately linear. The choice of initial displacements $d^{(2)}(\psi_1,\psi_2;t)$ admitted in Eq.~\ref{eq:d2} is constrained by the Hamiltonian structure of the GPE with its corresponding energy shells, i.~e.~$H[\psi_1]=H[\psi_2]=E$. With this constraint, the evolution of $\psi_1$, $\psi_2$ proceeds on a hypersphere in Hilbert space of fixed $E$. \\
It is instructive to first explore the short-time behavior of $d^{(2)}(\psi_1,\psi_2;t)$. Point of reference is the LSE, for which $d^{(2)}$ is strictly conserved, $d^{(2)}(\psi_1,\psi_2;t)=d^{(2)}(\psi_1,\psi_2;0)$. Variation of $d^{(2)}$ is therefore a direct measure for the influence of the non-linearity. For the non-linear GPE, we obtain for initial real wavefunctions $\psi_1(x,0)$ and $\psi_2(x,0)$ up to second order in t
\begin{equation}
\label{eq:re}
\operatorname{Re}{\langle\psi_1(t)|\psi_2(t)\rangle}=c_0+c_2t^2+O(t^4)
\end{equation}
with 
\begin{equation}
c_0=\langle \psi_1(0)|\psi_2(0)\rangle=1-d^{(2)}(\psi_1,\psi_2;0)
\end{equation}
and
\begin{eqnarray}
c_2&=&\frac{g}{2}\int {\rm d}x\;\psi_1(x,0)[n_1(x,0)-n_2(x,0),H_{\rm L}]\psi_2(x,0)\nonumber \\
&-&\frac{g^2}{2}\int {\rm d}x\;\psi_1(x,0)(n_1(x,0)-n_2(x,0))^2 \psi_2(x,0)\nonumber \\
\label{eq:c2}
\end{eqnarray}
with $n_i(x,0)=|\psi_i(x,0)|^2$ the condensate density. The square bracket in the first term of Eq.~\ref{eq:c2} denotes the usual quantum mechanical commutator. For strongly interacting condensates where the interaction energy is large compared to the single-particle energies, the $g^2$ term in Eq.~\ref{eq:c2} dominates. The initial quadratic increase of the distance function $d^{(2)}$ with time is universal and independent of integrability, non-integrability, or wave chaos. The short-time behavior of the distance function for a freely expanding condensate ($U(x)=0$) released from a harmonic trap displays the growth with $t^2$ followed by saturation (Fig.~\ref{fig:integ_d2}). We conjecture that this saturation is a signature of the integrability of the GPE for $U(x)=0$.\\
A slightly different problem, expansion of a condensate in zero potential ($U(x)=0$) inside a box defined by Dirichlet boundary conditions, was investigated by Villain and Lewenstein.\cite{VilLew00} Motivated by the structural similarity to the famous Fermi-Pasta-Ulam-Tsingou model, they observed stochastic behavior as extracted from a normalized statistical entropy. For sufficiently strong non-linearity a trend toward equipartition of energy among the modes of the LSE was found. The authors identified the free GPE with Dirichlet boundary conditions as non-integrable. We have investigated the behavior of the distance function $d^{(2)}$ for this system. We follow Ref.~\onlinecite{VilLew00} and chose as an initial state a Gaussian wavepacket $\psi_{\rm G}(x)$ with width $\sigma=1$ in a box of length $L=20$. This initial state is slightly distorted by a small linear admixture of strength $\alpha$:
\begin{equation}
\psi_{1,2}(x,0)=N_{1,2}(1\pm\alpha x)\psi_{\rm G}(x)
\label{eq:init_wfn_gauss}
\end{equation}
with normalization constants $N_{1,2}$. We test the sensitivity of $d^{(2)}(t)$ to $\alpha$ [i.e. to $d^{(2)}(0)$] by choosing two different values for $\alpha$: $\alpha=5\times10^{-4}$ which gives a similar initial $d^{(2)}(0)$ as in Fig.~\ref{fig:integ_d2} and $\alpha=5\times10^{-5}$ for which $d^{(2)}(0)$ is two orders of magnitude smaller. The numerical results confirm the expectation that $d^{(2)}(t)$ is linearly proportional to the initial distance $d^{(2)}(0)$ (constant shift in a logarithmic plot, see Fig.~\ref{fig:box_d2}). For the strength of the non-linearity we consider both a moderately strong value $g/L\approx79\epsilon_1$ where $\epsilon_1=\pi^2/2L^2$ is the ground-state energy of the LSE in the box comparable to the value ($g/L\approx61\epsilon_1$) in Ref.~\onlinecite{VilLew00}, and a much stronger non-linearity ($g/L\approx 1582\epsilon_1$) corresponding to $g_0$ (Eq.~\ref{eq:g0}). The energy unit in Ref.~\onlinecite{VilLew00} is related to our energy unit as $\epsilon_1=0.0123\hbar\omega_l$. The time unit $T_1=4L^2/\pi\approx500 t_0$ is long compared to our ``oscillation'' time scale $t_0$ characterizing the initial state.\\
For both non-linearities the distance function $d^{(2)}(t)$ increases quadratically with a slope approximately proportional to the non-linearity (see Fig.~\ref{fig:box_d2}) in qualitative accord to the short-time behavior (Eq.~\ref{eq:d2_2} and Eq.~\ref{eq:re}), however with much reduced slope. The growth seen in Fig.~\ref{fig:box_d2} proceeds slowly. The quadratic increase is fundamentally different from the exponential increase observed in the following sections.
This allows identifying within non-integrable systems those that do not exhibit wave chaos, as in Ref.~\onlinecite{VilLew00}, and those that do, as the systems discussed in the following.\\
The overall quadratic rise is modulated by fluctuations which result from partial revivals of the underlying LSE (see inset of Fig.~\ref{fig:box_d2}).
One prominent oscillation period is $T_1/8$ corresponding to the inverse energy spacing between the two lowest LSE states of even parity.\cite{VilLew00} Another prominent period $T_1/4$ results from the classical periodic orbit in the box. The non-linearity leads to a proliferation of frequencies by side-band coupling.\\
The important conclusion is that the non-integrable dynamics of the free GPE in a box with hard walls is, despite its complex appearance and the trend toward equipartition among the linear modes, non-chaotic as measured by the exponential separation of nearby trajectories in Hilbert space. Chaos requires the presence of non-vanishing external potentials $U(x)$.
\section{Numerical examples for wave chaos}\label{sec:examp}
We present in this section examples for expansion of BECs in potentials $U(x)$ of experimental relevance. For the generation of nearby initial states we use the following scenario: The BEC is created in a harmonic trap, i.~e.~the condensate wavefunction is given by the groundstate of the GPE, $\psi_{\rm g}(x)$, which is close to the Thomas-Fermi profile for the parameters chosen (see Sec.~\ref{sec:gpe}). Two nearby normalized states are then created by weak perturbations linear in x,
\begin{equation}
\label{eq:init_wfn}
\psi_{1,2}(x,0)=N_{1,2}(1\pm \alpha x)\psi_{\rm g}(x),
\end{equation}
with normalization constants $N_{1,2}$. Simultaneously with the release at $t=0$ the external potential $U(x)$ is switched on. For the perturbation parameter $\alpha$ controlling the distance between $\psi_1$ and $\psi_2$ we typically choose $\alpha=10^{-4}$. We have verified that the extracted values for the Lyapunov exponent $\lambda$ are independent of the value of $\alpha$.
\subsection{Weak periodic potential}\label{subsec:wpp}
We first consider the expansion of the BEC in weak periodic potentials
\begin{equation}
U(x)=U_0\cos{(2\pi x/l)}
\label{eq:pp}
\end{equation}
with periodicity $l$ and strength of the weak potential $U_0$. We require 
\begin{equation}
U_0/\epsilon\ll1,
\end{equation}
where $\epsilon=E/N$ is the total energy per particle of the interacting system. We will use in the following $U_0=0.2\epsilon$. The chemical potential is to a good degree of accuracy $\mu(t=0)=2\epsilon$ at $t=0$ for our choice of initial conditions. Unlike $\epsilon$, $\mu(t)$ is not conserved during the expansion and therefore, we use $\epsilon$ as characteristic energy scale. The periodic potential gives rise to a band structure of the linear problem with bandwidth of the first band, $W=\pi^2/2l^2$, (in literature also called recoil energy\cite{MorObe06}) and a band gap at the first Brillouin zone boundary, $\Delta\approx U_0$. The wide band gap limit corresponds to $W/U_0\ll1$, the nearly free particle limit to $W/U_0\gg1$. The spatial periodicity $l$ is assumed to exceed the coherence (healing) length $\xi$ of the condensate 
\begin{equation}
l\gtrsim\xi
\end{equation}
with
\begin{equation}
\xi=\frac{1}{\sqrt{8\epsilon}}.
\end{equation}
In the opposite limit $\xi\ll l$, the condensate can no longer resolve the local variation of the potential.\\
The initial wavefunctions are chosen as above (Eq.~\ref{eq:init_wfn}) having identical energy in the potential (Eq.~\ref{eq:pp}) and a Thomas-Fermi length of $L_{\rm TF}\approx8.4$ which is larger than the potential periodicity $l=5.8\xi$ with $\xi=0.0945$. The non-linearity is $g_0\approx390$. For $t>0$ the condensate wavefunctions expand and acquire an increasingly complex fluctuation pattern.
\begin{figure}
	\centering
		\includegraphics[width=8.5cm]{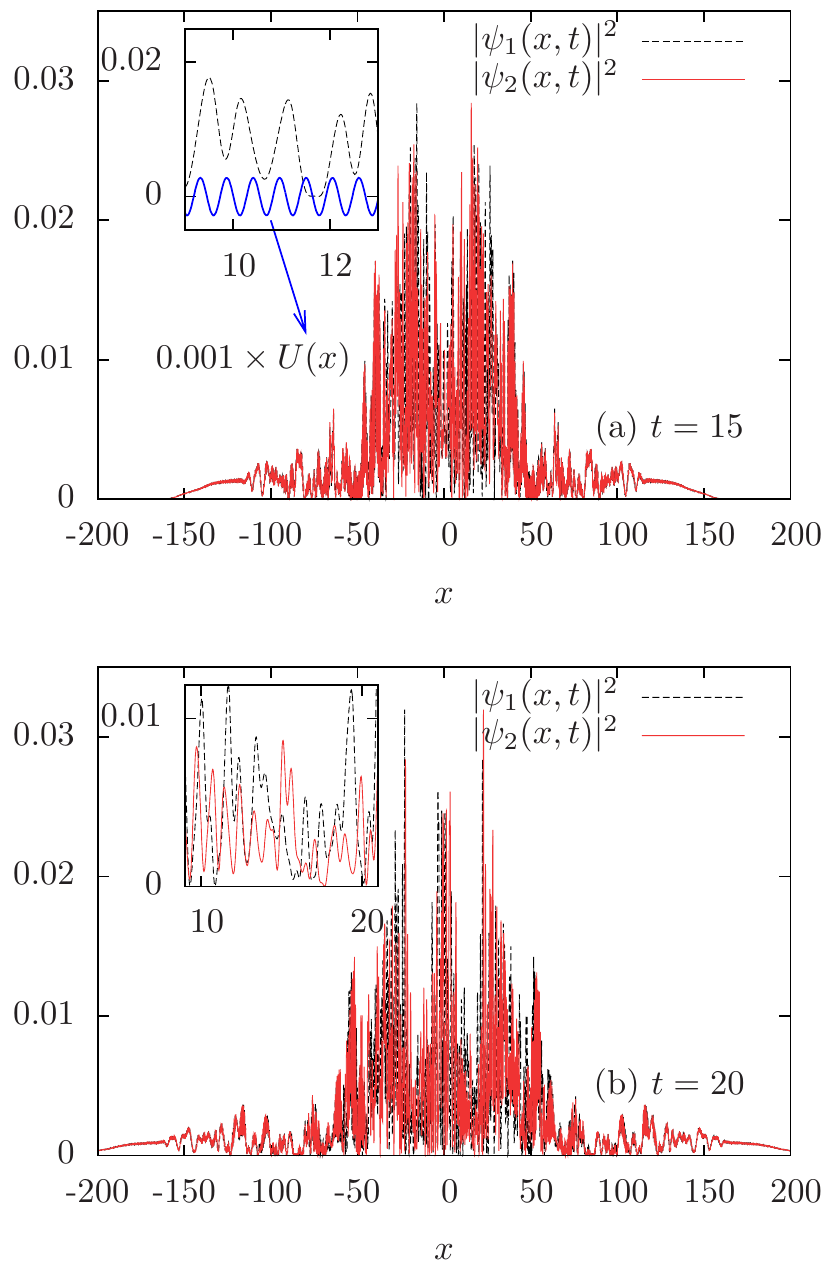}
	\caption{(Color online) Onset of divergence between $\psi_1$ and $\psi_2$ as a function of time. $|\psi_1|^2$ and $|\psi_2|^2$ are given for (a) $t=15$ and (b) $t=20$. Inset in (a): comparison of the spatial fluctuations of the wavefunction and the period of the potential $U(x)$. The potential parameters are $U_0=0.2\epsilon$ and $l=5.8\xi$. The non-linearity is $g_0$. Inset in (b): increasingly uncorrelated fluctuations of $|\psi_1|^2$ and $|\psi_2|^2$.}
	\label{fig:perlat_wfn}
\end{figure}
Appreciable deviations start to emerge near the center $x=0$ where fluctuations have the largest amplitude [see Fig.~\ref{fig:perlat_wfn} (a)]. The deviations increase with increasing time [see Fig.~\ref{fig:perlat_wfn} (b)], such that locally the amplitudes of $\psi_1$ and $\psi_2$ become uncorrelated [see the inset of Fig.~\ref{fig:perlat_wfn} (b)]. Such an extreme sensitivity to initial conditions is the hallmark of wave chaos. It is worth noting that on the length scale of $l$ both wavefunctions tend to mimic the oscillations of the potential while on larger length scales fluctuations of $\psi_1$ and $\psi_2$ lose their correlation. \\
We quantify the loss of correlation and the seemingly random fluctuations by the Lyapunov exponent $\lambda$ (Eq.~\ref{eq:lya}). Positive $\lambda$ signifies wave chaos. The time evolution of the distance function $d^{(2)}(\psi_1,\psi_2;t)$ resembles the time evolution of particle chaos. After an initial generic non-exponential (quadratic) growth (see Fig.~\ref{fig:perlat_d2_fb}), $d^{(2)}$ grows exponentially until the regime of fluctuations around the saturation value is reached. The slope of the exponential growth allows to numerically extract the Lyapunov exponent ($\lambda\approx0.7$ for the case of Fig.~\ref{fig:perlat_d2_fb}). 
\begin{figure}
	\centering
		\includegraphics[width=8.5cm]{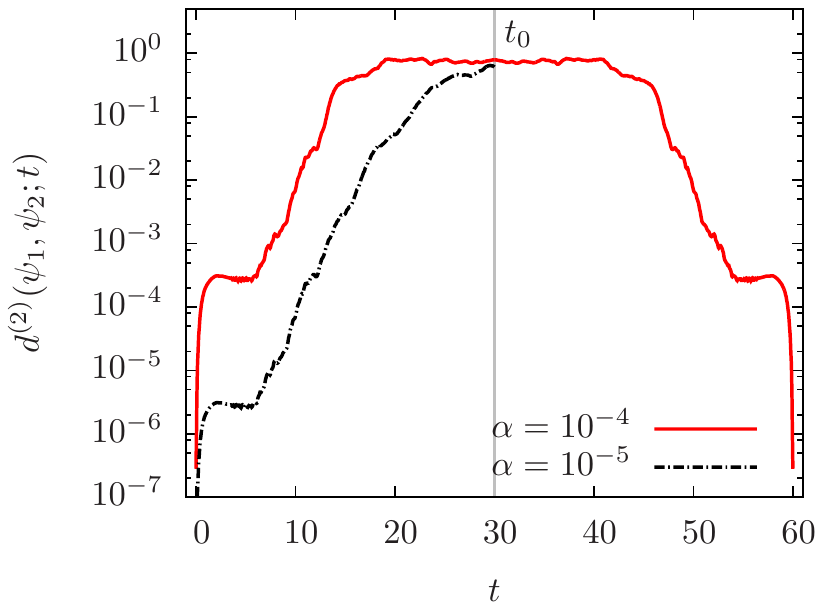}
	\caption{(Color online) Exponential growth of $d^{(2)}$ for two condensate wavefunctions propagated by the GPE in a weak periodic potential (same parameters as in Fig.~\ref{fig:perlat_wfn}). Time-reversed propagation for $t>t_0$ recovers the initial state precluding numerical noise as origin of the exponential growth. Two different distortion parameters have been used: $\alpha=10^{-4}$ and $\alpha=10^{-5}$.}
	\label{fig:perlat_d2_fb}
\end{figure}

\begin{figure}
	\centering
		\includegraphics[width=8.5cm]{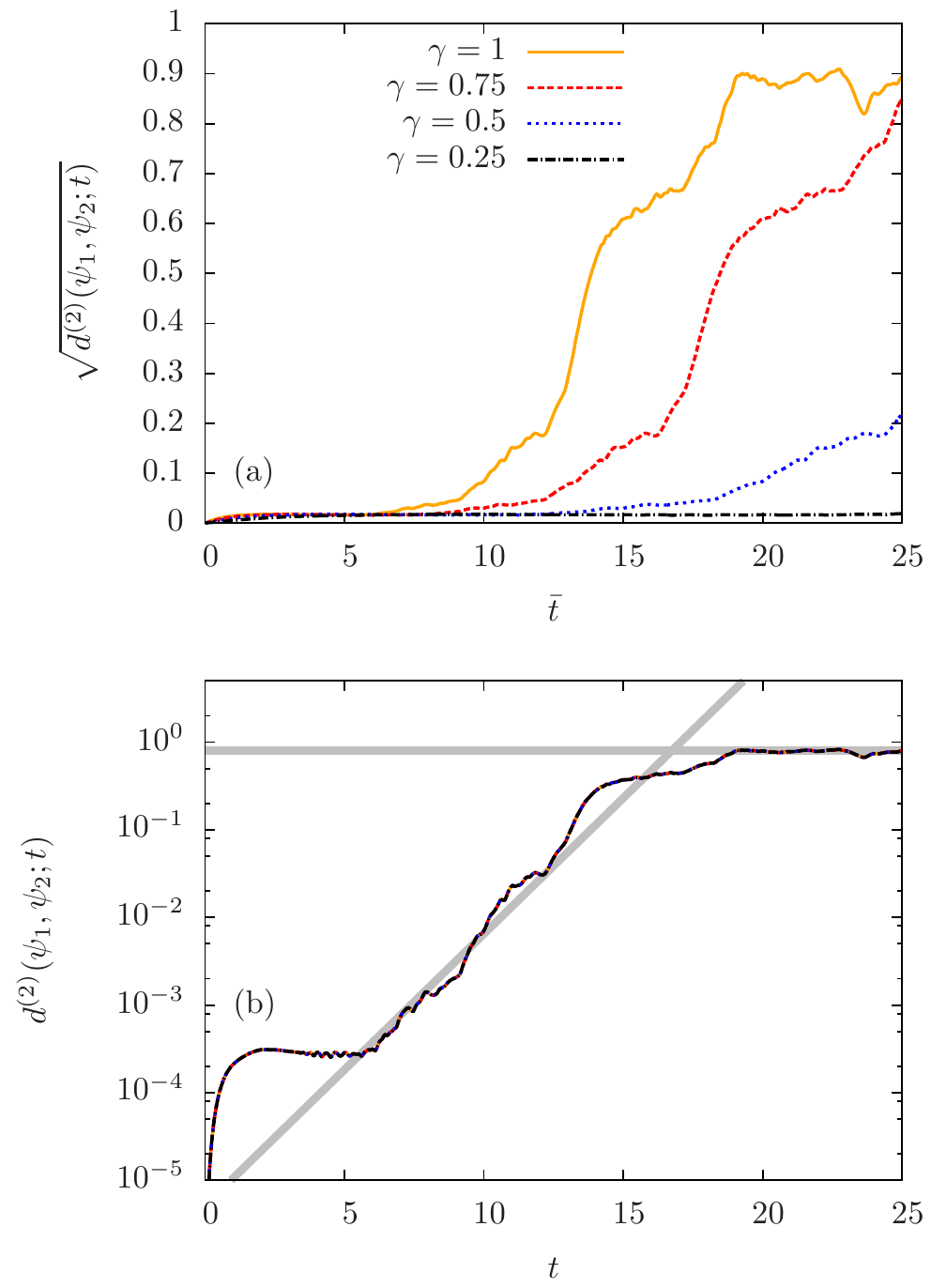}
	\caption{(Color online) Distance $d^{(2)}$ between $\psi_1$ and $\psi_2$ as a function of time for different $\gamma$. The parameters of the periodic potential are $\bar{U}=0.2\bar{\epsilon}$ and $\bar{l}=5.8\bar{\xi}$. (b) Logarithmic plot of (a) squared after rescaling time according to $t=\gamma \bar t$. The curves coincide. The gray solid lines underline the region of exponential growth and saturation and are guides for the eye.}
	\label{fig:perlat_d2}
\end{figure}

\begin{figure}
	\centering
		\includegraphics[width=8.5cm]{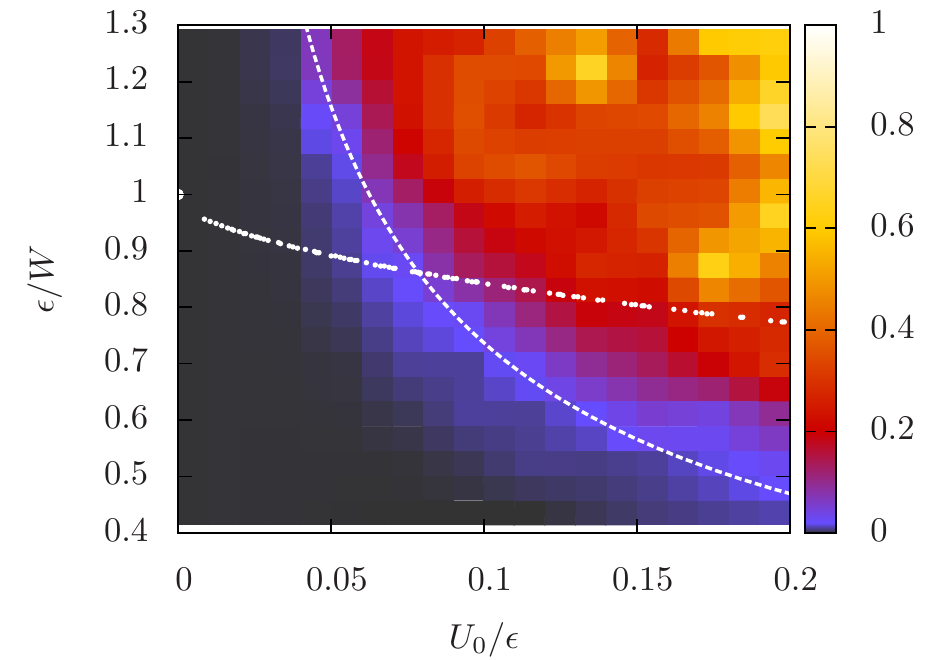}
	\caption{(Color online) The Lyapunov exponent $\lambda$ as a function of $\epsilon/W$ and $U_0/\epsilon$. $\epsilon$ is the energy per particle, $W=\pi^2/2l^2$ is the bandwidth of the first band of the linear system, and $U_0$ is the amplitude of the potential (Eq.~\ref{eq:pp}). The period $l$ is kept fixed $l=0.54811$. The same initial wavefunction has been used for all data points. $\lambda$ has been determined numerically via a linear fit to the logarithmic increase of $d^{(2)}$ (Eq.\ref{eq:d2}). The deviation (summed absolute difference between the linear fit and the numerical curve) is on average well below $0.4$ and maximal $\approx0.6$. To improve the graphical appearance every second point in the Figure is an interpolation between two numerical points. The white dashed curve is a fit to the threshold according to $\epsilon/W=0.165(U_0/\epsilon)^{-0.65}$. The white dots mark the point where the energy $\epsilon$ equals to the energy at the point of inflection of the first band.}
	\label{fig:perlat_lyap_trans}
\end{figure}

The deterministic nature of the the exponential separation can be verified by the time reversal test (Eq.~\ref{eq:bprop}). Upon reversal of the direction of time the two condensate ``trajectories'' approach each other again to distances close to their initial value (Fig.~\ref{fig:perlat_d2}). The distance function between $\psi_1(x,0)$ and $U_{-t}[\psi_1(x,t)]$ is of the order of $10^{-9}$. The near-perfect reversibility precludes numerical noise or ``numerical chaos'' as a source for divergence.\\ 
Furthermore, the scaling property of the GPE (Eq.~\ref{eq:gpe_rescal2}) is verified to a high degree of accuracy (Fig.~\ref{fig:perlat_d2}). We scale the initial conditions according to $\bar{\mu}(t=0)=\gamma\mu(t=0)$, such that $\bar U/\bar \epsilon$ is constant. Likewise the period of the potential is scaled such that $\bar l/\bar \xi=\sqrt{\gamma}l/\sqrt{\gamma}\xi$ is kept fixed. Despite a large absolute variation of $d^{(2)}$ with $\gamma$ [Fig.~\ref{fig:perlat_d2} (a)], upon rescaling, $d^{(2)}$ coalesces to a remarkable degree of accuracy [Fig.~\ref{fig:perlat_d2} (b)]. The $\gamma$ scaling is also equivalent to an effective change of the numerical grid in $(x,t)$, i.~e.~an increase of the numerical accuracy, since we have used the same absolute grid size for all calculations. The invariance of the results clearly demonstrates numerical stability. Since the Lyapunov exponent scales inversely with time $\bar \lambda=\gamma \lambda$, $\lambda$ decreases with decreasing $\gamma$ in accordance with Fig.~\ref{fig:perlat_d2} (a). Note, however, that this power-law scaling does not provide information on the existence (or absence) of critical thresholds for wave chaos (see below).\\
In the single-band mean-field Bose Hubbard model, a threshold for stochastic dynamics has recently been observed.\cite{CasMasDunOls09} We explore now the behavior of $\lambda$ in the continuum analogue, the GPE with a periodic potential. The Lyapunov exponent $\lambda$ will, in general, depend on the nonlinearity $g$, the period $l$ and the amplitude $U_0$ of the potential. The corresponding three energy scales are the total energy per particle $\epsilon$ (controlled by the non-linearity $g$) the bandwidth of the first band $W$, and the amplitude $U_0$. The scaling property allows to interrelate the dependences of $\lambda$ on these three parameters. For example one can investigate the dependence of $\lambda$ on the ratio $\epsilon/W$ at constant $U_0$ by keeping $\epsilon$ constant and decreasing $W$ (increasing $l$) giving the Lyapunov exponent $\lambda_{(\epsilon)}$, or keeping $W$ constant and increasing $g$, hence, $\epsilon$ giving $\lambda_{(W)}$. If the ratio $\epsilon/W$ remains invariant for the two cases, the Lyapunov exponents are interrelated by the scaling property (provided one of the initial wavefunctions is rescaled) as $\lambda_{(\epsilon)}=\gamma\lambda_{(W)}$ with $\gamma<1$. Therefore, it is sufficient to consider the ratio $\epsilon/W$ and not $W$ and $\epsilon$ separately. 
We arrive at a two-dimensional parameter plane ($U_0/\epsilon$, $\epsilon/W$). A scan of $\lambda$ is performed from the first to the second band of the linear system.\\
We observe a clear threshold of $\lambda$ in form of a rapid increase along both the $U_0/\epsilon$ and the $\epsilon/W$ axis (see Fig.~\ref{fig:perlat_lyap_trans}). Below the threshold $\lambda$ is vanishing. A power-law fit to the threshold line gives $\epsilon/W\propto(U_0/\epsilon)^{-0.65}$ (see dashed line in Fig.~\ref{fig:perlat_lyap_trans}). As the control parameter $\epsilon/W$ can be related to the parameter $\kappa$, the ratio between the non-linearity parameter and the hopping parameter of the discrete system, it is suggestive to assume that the transition in the discrete system\cite{CasMasDunOls09} and in the present continuous system are of common origin. However, the discrete system does not account for the transition along $U_0/\epsilon$. In other words, the mapping to the single-band Bose-Hubbard is not straightforward and a comparison to a multi-band model seems to be necessary.\\
Exponential divergence of wavefunctions above a threshold resembles the dynamical (or modulation) instability previously observed.\cite{WuNiu01,KonSal02,MorObe06} The modulation instability can be related to energetically surpassing the point of inflection along the first band of the linear problem: when the second derivative of the dispersion relation turns from positive to negative.\cite{KonSal02} The position of the inflection point, however, does not agree with the position of the threshold (in Fig.~\ref{fig:perlat_lyap_trans} marked by white dots). The reason may be that the simplified picture of modulation instability also does not account for the threshold along $U_0/\epsilon$. More importantly, the threshold does not seem to be sensitive to specific features of the linear problem, as presented in the next section. 
%
\subsection{Weak aperiodic potential}\label{subsec:wap}
%
\begin{figure}
	\centering
		\includegraphics[width=8.5cm]{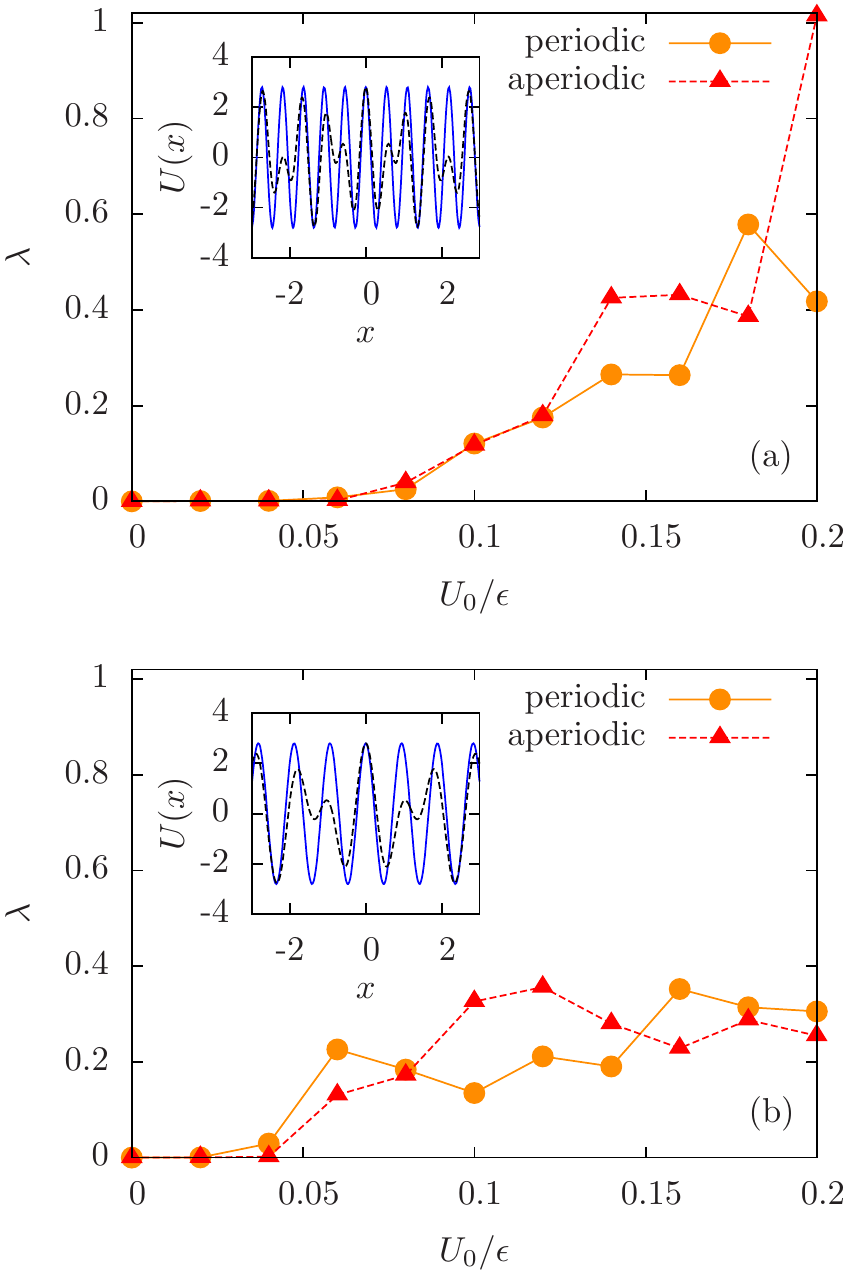}
	\caption{(Color online) Lyapunov exponent $\lambda$ as a function of $U_0/\epsilon$ for the periodic and aperiodic potential. The period of the periodic potential is (a) $l=0.54811$ and (b) $l=0.94503$. $l_1$ of the aperiodic potential (Eq.~\ref{eq:app}) is chosen respectively $l_1=l$. The potentials are given in the inset. The energy $\epsilon\approx14$ is constant such that $\epsilon/W$ is larger in (b) $\epsilon/W\approx2.5$ than in (a) $\epsilon/W\approx0.85$.}
	\label{fig:q_perlat_lyap_trans_small}
\end{figure}
In order to explore whether wave chaos in the GPE is a specific feature of the interplay between the periodic linear problem and the non-linearity, we investigate the dynamics in a strictly aperiodic potential
\begin{equation}
U(x)=U_0\left[c_1\cos{(2\pi x/l_1)}+c_2\cos{(2\pi x/l_2)}\right],
\label{eq:app}
\end{equation}
where $l_2/l_1=\frac{1}{2}(1+\sqrt{5})$ is the ``most'' irrational golden mean number and $c_1=c_2=1/2$. A well-defined band gap or well-defined negative mass dispersion relation near the Brillouin zone boundary which invoked the explanation of instability and stochastic motion\cite{WuNiu01,KonSal02,MorObe06} are absent for Eq.~\ref{eq:app}. For comparable values of $U_0/\epsilon$ as in the periodic case we find positive Lyapunov exponents (see Fig.~\ref{fig:q_perlat_lyap_trans_small}). Interestingly, the threshold as well as the overall behavior agree well with the periodic case, suggesting that chaos does not depend on the specifics of the linear system but rather on the overall length and energy scales. In agreement with Fig.~\ref{fig:perlat_lyap_trans} the threshold shifts toward smaller values of $U_0/\epsilon$ for higher ratios of $\epsilon/W$ [Fig.\ref{fig:q_perlat_lyap_trans_small} (a) and (b)]. Note that a characteristic energy $W$ can be defined for any potential with a characteristic length $l_c$ as $W\propto1/l_c^2$. Obviously, the wave chaos we have identified in the GPE is a general feature not coupled to specific properties of the bandstructure of the periodic linear problem.  
\subsection{Weak disorder}\label{subsec:wd}
\begin{figure}
	\centering
		\includegraphics{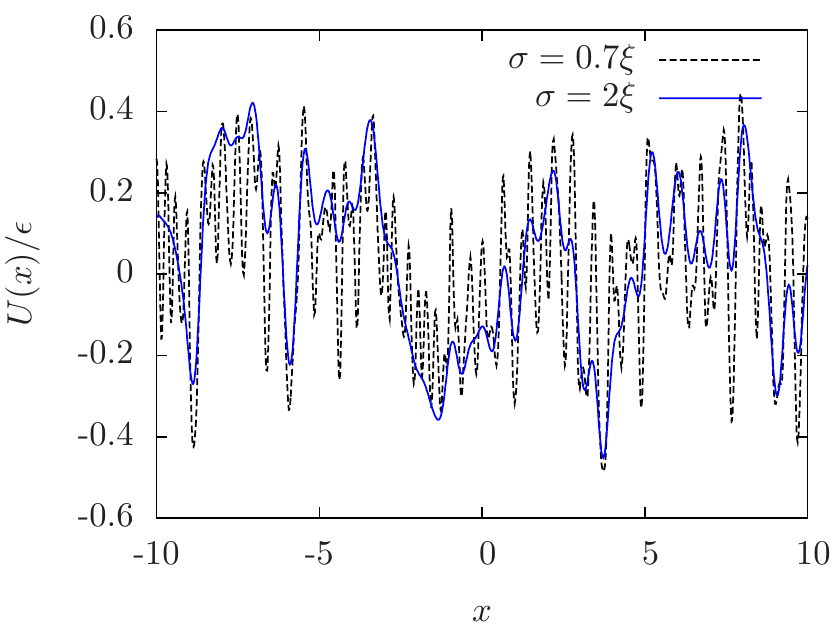}
	\caption{(Color online) Disorder potential for different values of the correlation length $\sigma$. The amplitude for both potentials is $U_0=0.2\epsilon$ and $N_d=2\times10^4$. The energy $\epsilon$ and healing length $\xi$ correspond to the groundstate of the GPE in a harmonic oscillator at non-linearity $g_0$ (Eq.~\ref{eq:g0}).}
	\label{fig:dis_pot}
\end{figure}

\begin{figure}
	\centering
		\includegraphics{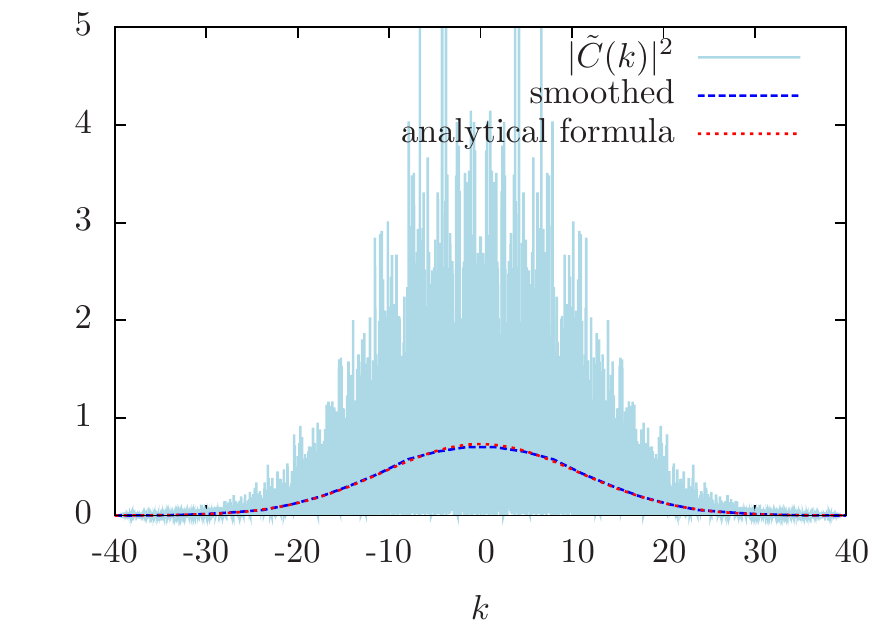}
	\caption{(Color online) The Fourier transform of the correlation function $\tilde C(k)$ for $U_0=0.2\epsilon$ and $\sigma=0.7\xi$ (Fig.~\ref{fig:dis_pot}). $\tilde C(k)$ agrees very well with Eq.~\ref{eq:dp_k} when smoothed over the fluctuations.}
	\label{fig:dis_pot_k}
\end{figure}

\begin{figure}
	\centering
		\includegraphics{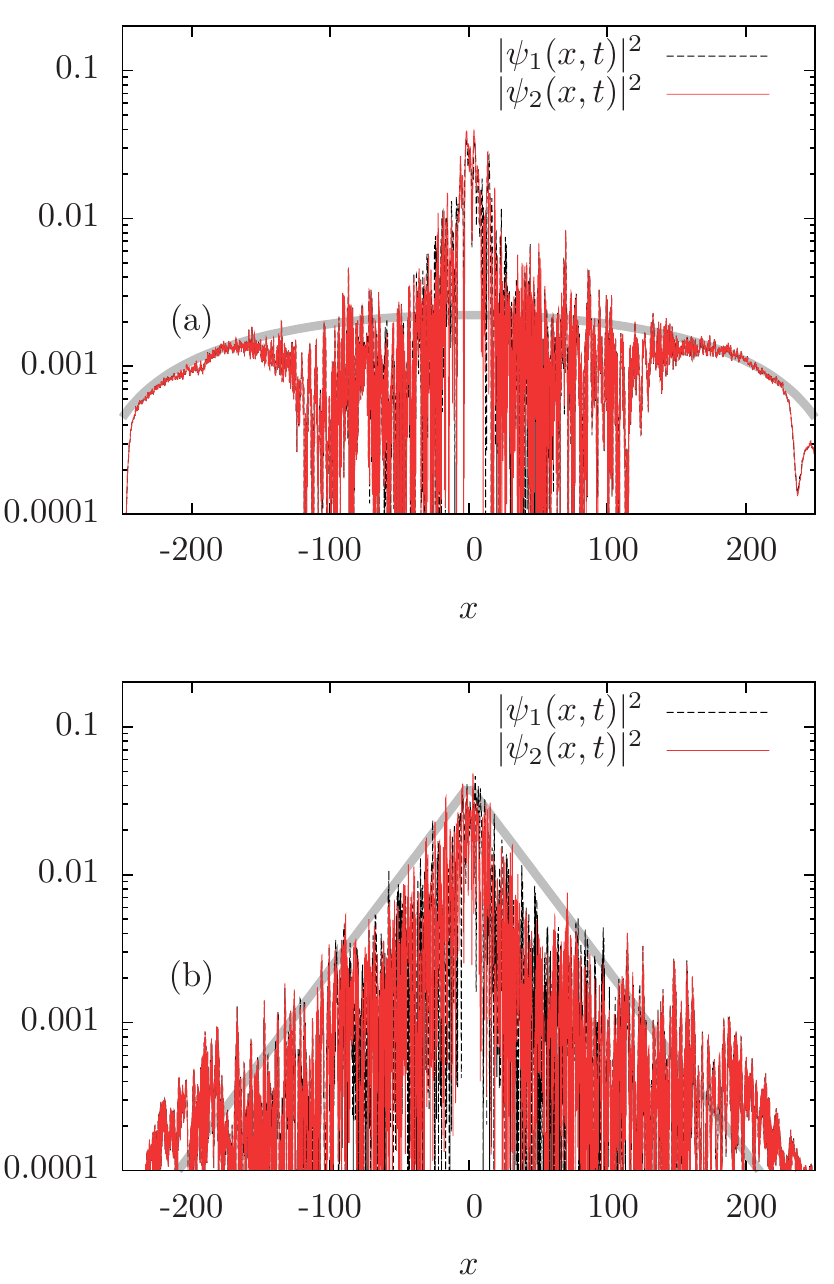}
	\caption{(Color online) Chaotic fluctuations of $\psi_1$ and $\psi_2$ at $t=24$ for (a) $\sigma=2\xi$ in the algebraic regime and (b) for $\sigma=0.7\xi$ in the ``localized'' regime (in the potentials of Fig.~\ref{fig:dis_pot}). The local fluctuations of $\psi_1$ and $\psi_2$ strongly deviate.}
	\label{fig:dis_wfn_sgn_sig07_2}
\end{figure}

A further generalization is the GPE with random disorder replacing the smooth (a)periodic potential. This case is of particular interest in the context of Anderson localization recently studied for expanding BECs.\cite{BilJosZuoCleSanBouyAsp08,Inguscio08} The latter experimental observations refer to the (quasi) linear regime when the non-linearity was sufficiently weak such that the localization dynamics is governed by the Schr\"odinger equation. Localization in the presence of interactions has remained an open question. For the discrete NLSE, subdiffusive expansion rather than localization was observed numerically.\cite{PikShe08,FlaKriSko09,LapBodKriSkoFla10} We focus on the interplay between the randomness induced by the potential and by the wave chaos in the GPE.\\
We follow closely the scenario employed in the investigations of Anderson localization:\cite{SanCleLugBouShlAsp07,BilJosZuoCleSanBouyAsp08} the BEC initially trapped in a harmonic oscillator expands in a disorder potential. We construct a disorder potential with Gaussian correlation and zero mean. At $N_{\rm d}$ equidistant grid points $x_i$ we place Gaussians with width $\sigma$ and random weight $A_i$
\begin{equation}
\tilde{U}(x)=\sum_{i=1}^{N_{\rm d}}A_i e^{\frac{-(x-x_i)^2}{2\sigma^2}}.
\end{equation}
After subtracting the mean, $\Delta U(x)=\tilde U(x)-\langle \tilde U(x)\rangle$ and normalizing by the standard deviation, $\langle \Delta U(x)^2\rangle^{1/2}$, the disorder potential becomes (Fig.~\ref{fig:dis_pot})
\begin{equation}
U_{\rm d}(x)=\frac{U_0}{\langle \Delta U(x)^2\rangle^{1/2}}\Delta U(x).
\label{eq:dp}
\end{equation}
For $U_0$ we choose the same values as in the cases of periodic and aperiodic potentials above. The disorder potential Eq.~\ref{eq:dp} is by construction Gaussian correlated, 
\begin{equation}
C(x)=\langle U_d(x_0+x)U_d(x_0) \rangle=U_0^2e^{\frac{-x^2}{(2\sigma)^2}}
\end{equation}
with a Gaussian shaped Fourier transform (Fig.~\ref{fig:dis_pot_k})
\begin{equation}
\tilde{C}(k)=\sqrt{2}\sigma U_0^2e^{-k^2\sigma^2}.
\label{eq:dp_k}
\end{equation}
Unlike for speckle potentials, the averages over odd powers $\langle U_d^{2n+1}\rangle$ vanish (numerically only approximately). The correlation length of the disorder potential $\sigma$ provides a length scale competing with the healing length $\xi$ of the free condensate.\\
For $\sigma<\xi$ the wavefunction is exponentially localized\cite{SanCleLugBouShlAsp07} which is characteristic for Anderson localization [see Fig.~\ref{fig:dis_wfn_sgn_sig07_2} (b)]. For $\sigma>\xi$ (we have chosen $\sigma=2\xi$), the condensate wavefunction develops algebraically decaying tails\cite{SanCleLugBouShlAsp07} that travel seemingly undisturbed through the disorder potential similar to propagation in free space [see Fig.~\ref{fig:dis_wfn_sgn_sig07_2} (a)]. 

\begin{figure}
	\centering
		\includegraphics{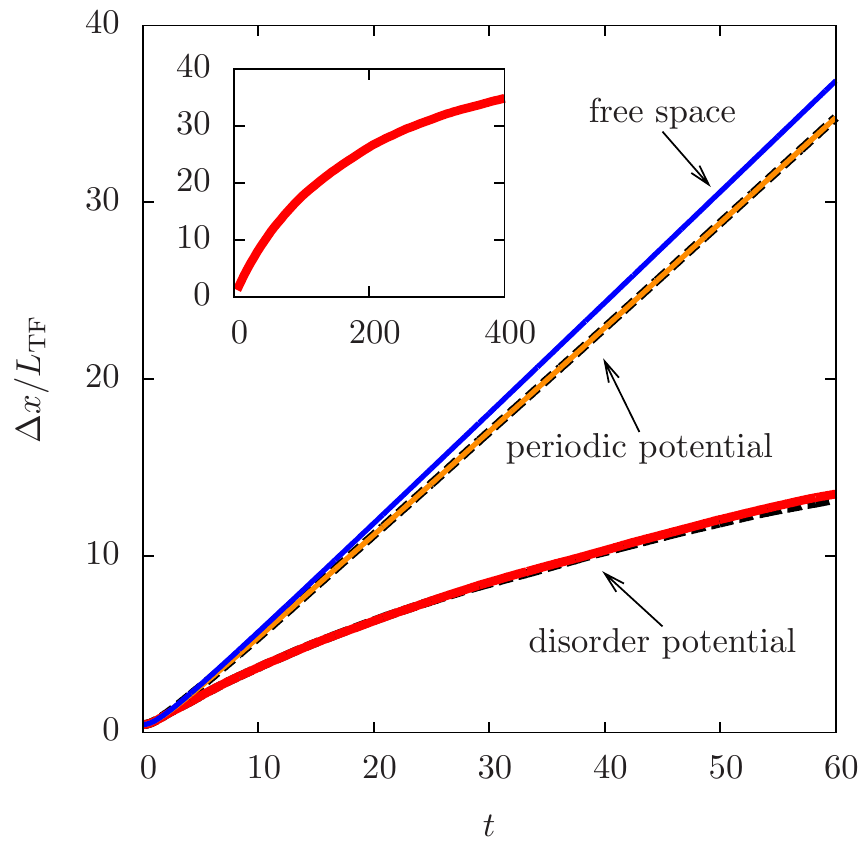}
	\caption{(Color online) The variance $\Delta x$ of the wavefunction $\psi_1(x,t)$ for propagations in free space, in a periodic potential ($U_0=0.2\epsilon$, $l=20\xi$) and in a disorder potential ($U_0=0.2\epsilon$ and $\sigma=0.7\xi$ as in Fig.~\ref{fig:dis_pot}). The black dashed lines correspond to $\psi_2(x,t)$ for the periodic and disorder potentials where chaos is present. The non-linearity is $g_0\approx390$. The inset gives a long time evolution for a weaker non-linearity $g_0/10$ in a disorder potential with $U_0=0.2\epsilon$ and $\sigma=0.32\xi$. In this case the initial wavefunction corresponds to the groundstate of the GPE for $g_0/10$.}
	\label{fig:delx}
\end{figure}

We probe for wave chaos for these two cases. It should be noted that in the presence of disorder, the determination of the exponential separation of nearby initial wavefunctions on the energy hypersphere faces the additional difficulty that it is not straightforward to create two nearby wavefunctions $\psi_1(x,0)$ and $\psi_2(x,0)$ with equal energy. In the present case the energies of the wavefunctions are kept close with a relative deviation of the order of $10^{-4}$.\\
In both cases, the algebraic and the exponential regime, we observe an exponentially growing separation and eventual saturation near the maximal value of separation $d^{(2)}(\psi_1,\psi_2)\approx 1$, confirming the intuitive notion that a random potential induces wave chaos. The Lyapunov exponent is $\lambda\approx0.85$ in the algebraic regime [Fig.~\ref{fig:dis_wfn_sgn_sig07_2} (a)] and $\lambda\approx0.86$ in the exponential regime [Fig.~\ref{fig:dis_wfn_sgn_sig07_2} (b)].\\
It is instructive to inquire into the relation between wave chaos and localization. Wave chaos is a signature of the non-linearity in the wave equation, while localization is the hallmark of disorder in the linear wave equation. In discrete non-linear models a variety of relations have been observed, one of which is subdiffusive growth rather than localization (see e.~g.~Ref.~\onlinecite{LapBodKriSkoFla10} and references therein). As measure for the localization dynamics we use the time evolution of the variance of the condensate wavefunction, $\Delta x=\sqrt{\langle x^2\rangle-\langle x\rangle^2}$. Note that our measure for the width is the square root of the second moment $m_2$ in Ref.~\onlinecite{LapBodKriSkoFla10}. We observe (Fig.~\ref{fig:delx}) for the expansion in a disorder potential a slowing down of the spread which locally scales like $\Delta x\propto t^{a}$. In qualitative agreement with the prediction in Ref.~\onlinecite{LapBodKriSkoFla10} the exponent is not constant but changes over time. We observe for the system in Fig.~\ref{fig:delx} with non-linearity $g_0$ an exponent of $a\approx0.75$ up to $t=60$ (compare our data also with Ref.~\onlinecite{SanCleLugBouShlAsp07}). To stay numerically on the safe side during long-time evolutions we performed also calculations with $g_0/10$. The exponent is predicted to be independent of non-linearity.\cite{LapBodKriSkoFla10} Note that for a direct comparison the time in units used in this manuscript is to be scaled by a factor\footnote{A direct conversion of units to these of Ref.~\onlinecite{LapBodKriSkoFla10} faces the difficulty that we do not use a finite difference grid but a FEDVR basis with non-equidistant grid points in space. For the conversion factor we have taken the mean grid spacing.} $1250$ to obtain the units of Ref.~\onlinecite{LapBodKriSkoFla10}. We observe that the exponent $a$ changes from $0.78$ between $t=0-100$ to $0.52$ between $t=100-300$ and becomes $0.34$ between $t=300-400$ in qualitative agreement to Ref.~\onlinecite{LapBodKriSkoFla10}. It should be noted, however, that the extraction of this exponent for the chaotic GPE is numerically much more challenging and less accurate than for discrete models. For example, the rigorous convergence test for time-reversed propagation (Eq.~\ref{eq:bprop}) begins to fail for the longest propagation times displayed in Fig.~\ref{fig:delx}.\\
Without spatial confinement of the system the density may become eventually sufficiently low such that the non-linearity ($g|\psi|^2$) can be neglected compared to the disorder potential and the system reacts by localization in the linear regime. The experimentally observed Anderson localization of expanding BECs\cite{BilJosZuoCleSanBouyAsp08} is likely associated with the approach of the linear regime. However, the presence of subdiffusive expansion due to residual non-linearities cannot be ruled out. \\
A further interesting observation is that the variance ($\Delta x$) for the two nearby initial states $\psi_1$ and $\psi_2$ agrees well (Fig.~\ref{fig:delx}). Despite the intrinsic randomness of the propagated condensate wavefunctions on length scales of the potential, do averages such as $\Delta x$ agree very well with each other.
%
\section{Discussion and conclusions}\label{sec:dis}
We have shown that the expansion of the condensate wavefunction described by the GPE features wave chaos as determined by a positive Lyapunov exponent in Hilbert space. We find wave chaos to be generic, i.~e.~present for a variety of one-body potentials $U(x)$ including weak periodic, aperiodic, and random potentials. The wavefunctions develop on the length scale of the healing length and of the local variation of $U(x)$ fluctuations with amplitudes and phases exponentially sensitive to the initial state such that nearby states become approximately orthogonal to each other. We note a remarkable exception: for harmonic potentials, $U(x)=\frac{1}{2}x^2$, wave chaos is absent as a consequence of the harmonic potential theorem for many-body states.\cite{Dob94}\\
The physical implications of these findings are of considerable interest and raise many conceptual questions. Wave chaos is a signature of the development of condensate density fluctuations. For a weakly interacting Bose gas condensate density fluctuations coincide with total density fluctuations of the many-body system and represent long-wave phonon excitations. For stronger interactions, the delayed onset of exponential divergence may delimit the characteristic time over which the GPE for the condensate is capable of describing the expanding BEC. Beyond this time, the expanding condensate is depleted by multiple excitations which the GPE attempts to mimic within a mean-field one-particle wavefunction by random fluctuations. Simply put, in this regime wave chaos may mark the breakdown of the mean-field approximation. Such reasoning would be based on the lack of exponential sensitivity to initial conditions in the real many-body system. If we consider two many-body wavefunctions for $N$ particles, $\psi_1(x_1,\ldots,x_N,t=0)$ and $\psi_2(x_1,\ldots,x_N,t=0)$, expanded in the basis of energy eigenfunctions $\psi_n(x_1,\ldots,x_N)$ of the underlying Hamiltonian: $\psi_1(x_1,\ldots,x_N,t=0)=\sum_n c_n \psi_n(x_1,\ldots,x_N)$ and $\psi_2(x_1,\ldots,x_N,t=0)=\sum_n c'_n \psi_n(x_1,\ldots,x_N)$ the distance function $d^{(2)}(\psi_1,\psi_2)$ remains constant irrespective of the choice of the underlying Hamiltonian. The overlap of wavefunctions does not change in time. This fundamental property of linear time evolution should be reproduced by any approximative theory. However, the difficulty with this argument lies in the fact that the distance function $d^{(2)}$ in the many-body Hilbert space $\mathcal{H}=\prod\nolimits_{i=1}^N (L^2)^i$ is unrelated to that in the $L^2$ space of the reduced mean-field wavefunction entering the GPE. Moreover, ensemble expectation values, such as the moments of the distribution (e.~g.~$\Delta x$ discussed above) which represent averages over the condensate wavefunction over fine-scale density fluctuations, are found to be insensitive to small variations of the initial state. The GPE may remain predictive for such expectation values on longer times scales. The latter would explain the large number of successful applications of the GPE to condensate expansions (see e.g. Ref.~\onlinecite{LyeFalModWieForIng05,BilJosZuoCleSanBouyAsp08,Inguscio08,CheHitJunWelHul08,DriPolHitHul10}). Such a link between chaotic wavefunctions and observables would be analogous to classical chaos for particles and phase space distribution functions: While the long-time evolution of individual trajectories become unpredictable, the ensemble or time average over stochastic regions in phase space remain well defined and yield stable ensemble expectations values. Going beyond large-scale averaging, the characterization of the fine-scale random fluctuations developing under propagation with the GPE remains a widely open problem. Propagation of many-body wavefunctions beyond the GPE may shed light on the question to which extent the GPE fluctuations, in the mean-field level, ´may faithfully represent at least some of the features of multi-particle excitation. The latter may also provide insight into the applicability of a wavefunction ``thermalization hypothesis'' originally developed for wavefunctions of the LSE for classically chaotic systems\cite{Sre94} to wave chaos in the GPE. Furthermore, it would be of interest to experimentally search for and characterize local fluctuations and structures in expanding condensates, e.~g.~by elastic light scattering.
\section*{Acknowledgments}
We thank Julien Armijo, Denis Basko, Philippe Bouyer, Moritz Hiller, Ansgar J\"ungel, Daniel Matthes, and Markus Oberthaler for helpful discussions and hints on literature. We further thank Dan Horner, Florian Libisch, Stefan Nagele, Renate Pazourek, and especially Johannes Feist for providing valuable input on the numerical part of this work. This work was supported by the FWF program ``CoQuS'' and by FWF-SFB 041 ``ViCoM''. Calculations have been performed on the Vienna Scientific Cluster and under Institutional Computing at Los Alamos National Laboratory on the Coyote and Lobo platforms. The Los Alamos Laboratory is operated by Los Alamos National Security, LLC for the National Nuclear Security Administration of the U.S.~Department of Energy under Contract No.~DE-AC52-06NA25396. 

\end{document}